# How do 'technical' design-choices made when building algorithmic decision-making tools for criminal justice authorities create constitutional dangers?

by Karen Yeung[α] and Adam Harkens[β]

## Part II

### 1. Introduction

The high-profile failure of automated digital decision tools by public authorities, particularly those that utilise some form of machine learning (ML), including the 'robo-debt' scandal in Australia[1] and the Dutch childcare benefits scandal,[2] illustrate the scale and seriousness of the hardship and injustice that digital 'solutions' in government can produce. Reflecting on these systems, the UN's previous Special Rapporteur on Extreme Poverty, Philip Alston has warned of the imminent dangers of 'digital dystopia'[3] highlighting the urgent need for safeguards. Although established public law principles *could* be mobilised to prevent mistakes and failures, they are yet to be effectively and systematically operationalised in the development, implementation, and oversight of public sector algorithmic tools. This two-part paper focuses on digital tools used by criminal justice authorities that purport assess the 'risk' posed by specific individuals to inform how they should be treated, although much of our analysis has wider applicability to the public sector more generally. In Part I, we showed how ML-based automated digital decision-making and support tools ('algorithmic tools') are conventionally built and implemented without reference to their larger context of application. We argued that, despite the 'rights-critical' nature of criminal justice decisions,[4] algorithmic tools for public sector use are conventionally developed in ignorance of public law principles and the legal duties to which they give rise. For technical developers, these contextual considerations are conventionally regarded as irrelevant 'noise', informed by what we call a 'contextual detachment mindset'.[5] As a result, vital institutional safeguards against the arbitrary or otherwise unjust exercise of power by public authorities are being circumvented, substantially enhancing the likelihood that public powers may be exercised unlawfully, creating injustice that is in practice difficult to detect and almost impossible to challenge. In this Part II, we demonstrate more precisely how choices made by technical developers during the design, construction and implementation of algorithmic tools implicate several legal duties that apply to the exercise of the decision-making authority that those tools inform, including English administrative law doctrine, human rights protections set out in the European Convention on Human Rights

---

[α] Professorial Fellow in Law, Ethics and Informatics, Birmingham Law School and School of Computer Science, University of Birmingham. We gratefully acknowledge funding support form VW Stiftung, Grant No: 19-0087 (2019-2023) and for helpful feedback by Emma Ahmed-Rengers (particularly in comparing conventional statistics with ML approaches), Reuben Binns, Mireille Hildebrandt, Tobias Krafft, Winston Maxwell, Leandro Minku, Johannes Schmees, Georg Wenzelberger and Katharina Zweig. Karen Yeung drafted the text and devised the analytical framework, argument, paper structure and narrative. Adam Harkens undertook the in-depth case-studies and background research to the legal, scholarly and contextual detail supporting Yeung's arguments and acted as a critical sounding board for her ideas. An earlier version was presented by Karen Yeung to the Norwegian Association for Computers and the Law, The Knut Selmer Memorial Lecture, 23 November 2020 (Oslo).
[β] Post-doctoral Research Fellow, Birmingham Law School.
[1] P. Henman, "Administrative justice in a digital world: Challenges and solutions" in J. Tomlinson, R. Thomas, M. Hertogh and R. Kirkham (eds.) *The Oxford Handbook of Administrative Justice* (Oxford: OUP, 2021).
[2] Melissa Heikilla, "Dutch scandal serves as a warning for Europe over risks of using algorithms" (29 March 2022, *Politico EU*), https://www.politico.eu/article/dutch-scandal-serves-as-a-warning-for-europe-over-risks-of-using-algorithms.
[3] Philip Alston, 'Report of the Special Rapporteur on Extreme Poverty and Human Rights' A/74/48037 (New York: United Nations, 2019)**.**
[4] i.e., such decisions may interfere with the legal and fundamental rights of affected individuals: see Part I section 3.
[5] See K.Yeung and A. Harkens, "How do 'technical' design-choices made when building algorithmic decision-making tools for criminal justice authorities create constitutional dangers? Part I" [2022] *Public Law*, forthcoming at section 2.2.

(ECHR) and incorporated by the Human Rights Act (HRA), data protection laws arising under that Data Protection Act 2018, and the so-called Public Sector Equality Duty.[5] In so doing, we seek to move beyond existing academic inquiry that has tended to be rather general and abstract. This high-level of generality is problematic because, as legal scholars commissioned by the Administrative Conference of the United States (ACUS) to review federal agency use of algorithmic tools have observed:

> "…much, if not most, of the hard work regulating algorithmic governance tools will come not in the constitutional clouds, but rather in the streets of administrative law."[6]

We begin with a brief account of three algorithmic tools that purport to assess the 'risk' posed by individuals ('i-RATS') currently (or until recently) in use by criminal justice authorities: two tools used in England - the London Gangs Matrix and the Durham Constabulary's HART tool, and the SyRI tool formerly used in the Netherlands. We then examine the intersection between public law and data science perspectives by adopting the lens of 'algorithmic regulation', drawing selectively from these three i-RATs, to demonstrate how particular abstraction decisions involved in algorithmic model-building directly implicate constitutional principles at each stage of the development process yet are conventionally and routinely ignored. We argue that algorithmic tool-developers, and the authorities who commission and implement them, have failed to recognise, or understand, the constitutional and legal implications of these technical choices. Hence, algorithmic tools are being employed by criminal justice authorities in ways that unjustifiably violate constitutional principles and the specific legal duties to which they give rise, significantly enhancing the risk and magnitude of injustice and abuses of power that can arise from their use.

The third and fourth sections consider the implications of our analysis. We suggest that constitutional principles should be mandatory requirements forming an essential part of the design-brief which those who build algorithmic tools to inform criminal justice decision-making must adhere to. However, because technical developers cannot be expected to understand nor properly apply public law principles and duties, we argue that they must collaborate closely with legal experts when deciding whether to deploy these tools for specific criminal justice purposes, and if justified, to ensure that they are configured in a manner that is demonstrably compliant with public law principles (including respect for human rights) and all applicable legal duties throughout the tool-building process. If such compliance cannot be demonstrated, they should *not* be used. Given that lawyers unfamiliar with algorithmic model-building may struggle to understand how such principles and duties are implicated in technical design-choices, we also offer several practical recommendations, highlighting several 'detachment practices' (that is, practices in which legally and constitutionally relevant matters are conventionally ignored by technical developers) that must be *avoided* when algorithmic tools are developed for use by criminal justice decision-makers. Finally, we outline a series of urgently-needed systematic legal reforms to help establish and maintain public trust in criminal justice decision-making in an increasingly algorithmic age, followed by a brief conclusion.

2. **Three algorithmic risk assessment tools used to categorise and 'flag' individuals**

We begin with a brief account of three i-RATS used in criminal justice settings based on knowledge gleaned largely from publicly-available documents: The London Gangs Matrix, the Durham 'HART' tool, and the Dutch SyRI tool. Although their technical dimensions and intended purposes vary significantly (to reduce gang violence, to enhance the effectiveness of offender 'rehabilitation', and more efficient identification of social welfare fraudsters respectively), they all produce an algorithmically generated assessment of an individual's 'risk' for use by front-line decision-makers when deciding what action to take against those individuals. It is important to emphasise, however, that their effectiveness in achieving their claimed criminal justice purposes has *not* been systematically evaluated, let alone proven, and this, as we shall see, is a matter of considerable constitutional importance given that they entail prima face interferences with human rights.[8]

---

[5] Equality Act 2010, s.149.
[6] David Freeman Engstrom et al., "Government by algorithm: Artificial intelligence in federal administrative agencies" (Report submitted to the Administrative Conference of the United States, 2020), 76.
[8] E.g., consider the 'Waterproof' project referred to in the *SyRI* case, discussed in section 3.1 of Part I.

*2.1    The London Gangs Matrix*

The Gangs Matrix was created by the London Metropolitan Police Service (MPS) following the London riots in August 2011 to help reduce 'street-focused' gang violence. It provides police officers, via their internet-enabled smart devices, with a dynamic digital dashboard and database that lists individuals identified as potentially 'at risk' of involvement in gang violence (called 'Gang Nominals') together with a 'harm score' generated via an algorithmic assessment tool believed to be created using ML techniques.[9] The MPS claims that the Gangs Matrix (a) helps police officers identify and assess in real-time the 'risk of violent re-offending' of a Gang Nominal by classifying them as 'high' (red), 'medium' (amber) or 'low' (green); and (b) helps allocate MPS enforcement resources, prioritising those deemed most dangerous while others are diverted and offered alternative support. Individuals can be included on the Matrix if identified as a potential gang member by a police officer or partner agency representative (including local councils, schools, and health services) and this has been corroborated by 'reliable intelligence from more than one source.' Intelligence deemed 'reliable' could include the fact that the individual has been observed associating with, being related to, or subject to a stop and search report together with a Gang Nominal, or even sharing social media content that refers to a specific gang. Neither details of the model used to calculate the 'harm score' nor the data used to create it have been published, although the MPS states that factors contributing to the score include previous history of violence (weighted by seriousness of offence), violence or weapons intelligence in the previous six months, judgments of a local gang unit intelligence manager, and partner organisations' assessments. How these factors are weighted or utilised to calculate an individual's harm score is not publicly known: the MPS merely state that it is based on a 'complex scoring system'.[7] Individuals listed on the Matrix may be subjected to heightened police surveillance, often leading to pre-emptive stop and search, arrest for minor offences, anti-social behaviour injunctions and/or criminal behaviour orders. Although not originally intended, the Matrix is also allegedly used in evidence to support the prosecution of gang-related offences. Individuals listed on the Matrix have been subjected to police stop and search more frequently than the general population, with subsequently reduced levels of police attention for those whose details are removed. Individuals identified as potential 'victims' of gang crime and other vulnerable persons may also be included in the Matrix, albeit accompanied by a 'zero harm' score, as are individuals designated as Gang Nominals with no previous criminal convictions.

*2.2 The Harm Assessment Risk Tool (HART)*

HART is an algorithmic tool that, until recently,[8] was used by Durham Constabulary to help custody officers decide whether arrested individuals should be offered an opportunity to participate in its 'Checkpoint' rehabilitation program. HART was developed collaboratively by Durham Constabulary and Cambridge University researchers using 'custody event data'[9] drawn from the Constabulary's custody management IT systems for the five years to 31 December 2012.[10] Considerably more information about the HART tool is available than is typically the case for public sector decision-support tools, thanks to Sheena Urwin (then Head of Criminal Justice at Durham Constabulary), who wrote her Master's thesis under the supervision of Dr

---

[9] Amnesty International United Kingdom Section ("Amnesty International"), "Trapped in the Matrix: Secrecy, stigma, and bias in the Met's Gangs Database" (May 2018) 20 https://www.amnesty.org.uk/files/reports/Trapped%20in%20the%20Matrix%20Amnesty%20report.pdf, 13.

[7] MOPAC, "Review of the MPS Gangs Matrix" at https://www.london.gov.uk/sites/default/files/gangs_matrix_review_-_final.pdf, 20; Amnesty International, "Trapped in the Matrix," 11.

[8] House of Lords Justice and Home Affairs Committee, "Technology rules? The advent of new technologies in the justice system" (HL 2021-2022, 180-1).

[9] A 'custody event' refers to the disposal decision taken by the custody officer following arrest at the end of the first custody period, either to grant bail, remand D in custody, taken no further action, administer an out of court disposal, or prosecute the subject (with a decision to bail): Sheena Urwin, "Algorithmic forecasting of offender dangerousness for police custody officers: An assessment of accuracy for the Durham Constabulary model", Research Presented as for the purposes of gaining a Master's Degree in Applied Criminology and Police Management at Cambridge University (2016) at http://www.crim.cam.ac.uk/alumni/theses/Sheena%20Urwin%20Thesis%2012-12-2016.pdf, 37.

[10] Urwin, "Algorithmic forecasting of offender dangerousness," 37. See Teresa Scantamburlo, Andrew Charlesworth and Nello Cristianini, "Machine Decisions and Human Consequences," in K. Yeung and M. Lodge (eds.), *Algorithmic Regulation* (Oxford: Oxford University Press, 2019) which explains the technical development of the HART tool.

Geoffrey Barnes - one of the academics credited with designing HART.[11] Urwin's thesis,[12] which she published online, explains that HART used thirty-four risk predictors based on data about the arrested person at the time of arrest, combined with data from Durham Constabulary's pre-existing records, created using a random forest ML modelling technique. HART purported to calculate the arrestee's risk of committing an offence in the subsequent two years, described as a prediction of 'offender dangerousness' (a misnomer: those arrested have not been convicted and were therefore wrongly described as 'offenders'). Individuals predicted as likely to commit a 'serious' offence (defined as an offence involving violence), a non-serious offence, or no offence, were classified as 'high', 'moderate' or 'low' risk respectively.[13] Only those receiving a 'moderate' prediction were eligible for Checkpoint. Checkpoint participants are provided with an 'out-of-court disposal' which involves signing a 4-month deferred prosecution contract. Provided they meet the contract conditions throughout (including refraining from offending)[14] participants avoid formal prosecution and a criminal record. Individuals could not be diverted into Checkpoint unless the custody officer was satisfied that there was sufficient evidence to charge, in line with Crown Prosecution Service requirements.

*2.3    SyRI*

SyRI was an ML-based decision-support tool used by several Dutch public authorities to help detect tax and social security fraud until it was declared unlawful following a successful judicial review challenge.[15] SyRI was built from data extracted from individual's administrative records held by government agencies to generate individual risk profiles to identify suspected welfare fraud (i.e. those with "an increased risk of irregularities" when compared to the target population).[16] Potential risk indicators were based on demographic data (e.g., name, address, and associated administrative data), administrative fines and sanctions, debt burdens, social benefit, and tax. If algorithmically flagged, the individual's profile was then investigated by the Ministry of Social Affairs and Employment. Little is publicly known about the substantive interventions that followed once individuals had been notified that they had been flagged for investigation, but reportedly include administrative sanctions/fines, mandatory participation in civic education programmes, or in sufficiently serious cases, referral to the Public Prosecution Service.[17] In short, SyRI offered Dutch authorities a powerful technological means for monitoring individuals in real-time, linking vast swathes of personal, economic and administrative data collected from records held by multiple governmental organisations to inform how public authorities exercise their investigative and enforcement discretion.

## 3.    Understanding tool-building through the lens of 'algorithmic regulation'

In Part I, we explained how statistical prediction models built using ML techniques conventionally entails making abstraction decisions that intentionally ignore context-specific features of the real-world domain in which the tool will operate. For algorithmic tools intended to inform criminal justice decision-making, ignoring these context-specific features includes ignoring the larger constitutional backdrop against which these decisions are made. We now demonstrate more precisely how particular design-choices in algorithmic tool-building implicate, and must be constrained by, public law principles and legal duties by adopting the lens of 'algorithmic regulation.'[18] This analytical approach begins from the perspective of the data scientist commissioned to build an algorithmic tool for a specific organisational purpose, tracing the steps and design choices involved and then proceeds to critically investigate the points of contact between the algorithmic tool and the social world of its deployment, particularly their impacts upon those subjected to algorithmic evaluation. Although this second stage allows many different analytical perspectives,[19] we focus on the constitutional principles that a commitment to constitutionalism would impose as non-negotiable requirements – or 'parameters' to use more familiar computer science terminology – that should condition those choices. By decomposing ML tool-building into four-steps, we demonstrate in concrete terms how design choices implicate constitutional principles and the specific legal duties that operate to inform and constrain those choices.

---

[11] Geoffrey Barnes was then an Affiliated Lecturer at Cambridge University's Institute of Criminology; see Josh Jacobs 'The radical idea to reduce crime by policing less, not more' (Wired, 10 March 2021) *https://www.wired.co.uk/article/evidence-based-policing*. Information about HART is available from an academic journal article jointly written by Barnes, who was responsible for its technical design, and Urwin, who occupied a senior role in the Durham Constabulary at the time of its use - along with legal scholars Oswald and Grace: Marion Oswald et al., "Algorithmic risk assessment policing models: lessons from the Durham HART model and 'Experimental' proportionality" (2018) 27(2) *Information & Communications Technology Law* 223-250. Although the paper claims to offer a 'critique' of the HART tool, we do not understand why the authors declared that there was "no potential conflict of interest" at 250.
[12] Urwin's thesis sought to assess the validated accuracy of the HART model: "Algorithmic forecasting of offender dangerousness." 37; Oswald et al, "Algorithmic risk assessment policing models," 229-230.
[13] Urwin, "Algorithmic forecasting of offender dangerousness," 15.
[14] Contracts can include up to five conditions: e.g., refraining from re-offending, participating in restorative processes with victims, attending mental health or substance abuse therapy sessions, or wearing a GPS tag: Kevin Weir, Gillian Routledge & Stephanie Kilili, "Checkpoint: An Innovative Programme to Navigate People Away from the Cycle of Reoffending: Implementation Phase Evaluation" (2019) 15 *Policing: A Journal of Policy and Practice* 1-19, 9.
[15] NJCM et al. and FNV, Rechtbank den Haag, ECLI: NL: RBDHA: 2020:1878 ("SyRI Judgment").
[16] E.g., the Tax and Customs Administration, municipal authorities, the Public Prosecution Service, the police, and the immigration service: SyRI judgment, [3.7].
[17] SyRI Judgment at [4.17]; [6.59].

For simplicity, we consider a specific set of decisions made under conditions of uncertainty involving some kind of classification task: such as distinguishing 'spam' emails from legitimate ('ham') emails, to identify whether a patient is a diabetic or not, or to evaluate whether a person currently serving a custodial sentence for a criminal offence will re-offend on release. Imagine an organisation whose front-line decision-makers are routinely required to make decisions of this kind, and commissions the building of an algorithmic tool to assist them. To create such a tool, the algorithmic model-building process proceeds in four steps[20]:

**Step 1:** Identify suitable input data to serve as 'ground truth' for model-building;

**Step 2:** Create an algorithmic model by applying ML techniques to generate predictions concerning future outcomes on unseen data;

**Step 3:** Test and validate the model; and

**Step 4:** Encode the model into a software program capable of communicating its outputs via a digital interface to create a convenient digital tool to assist front-line workers.

*3.1     Step 1: Obtain a suitable dataset to serve as ground-truth to train a prediction model*

First, the developer must identify a dataset to serve as 'ground truth' in which each data item possesses the relevant feature of interest upon which to train the model.[21] Properties used to describe the item to be classified are often called 'features' while the classes assigned to each item are called 'labels'.[22] The relevant features of any given item depend upon the phenomenon which the developer chooses to rely on as the basis for classification. For our three examples, the features might refer respectively to: words used in the email, a set of clinical indicators, or a person's criminal history. The labels applied to the training data comprised of these items might be respectively labelled 'spam' or 'ham', 'diabetic' or 'healthy', or 'safe' or 'dangerous.' A machine learning algorithm may then be applied to this labelled training data to build a mathematical function (a 'classifier') that assigns a class label (e.g., 'spam' or 'ham') to any object (e.g., emails, patients, prisoners) that has not yet been labelled.[23] The resulting classifier can be used to make predictions on unseen data, helping to inform (or even to automate) decisions for a specific organisational purpose, such as the creation of an automated email-filtering tool, a screening tool to help identify patients likely to be (or become) diabetic, or to assist a parole board deciding whether to grant a prisoner early-release.

---

[18] Karen Yeung, "Algorithmic regulation: A critical interrogation" (2018) 12 (4) *Regulation and Governance* 505-523.
[19] See Lena Ulbricht and Karen Yeung "Algorithmic regulation: A maturing concept for investigating regulation of and through algorithms" (2021) 16(1) *Regulation and Governance* 3-22 on the 'thin' nature of algorithmic regulation as an analytical frame and the 'thicker' variety of conceptual lenses it allows.
[20] For a basic explanation of how ML-based models utilising supervised learning are created, see Part I section 1.
[21] David Lehr and Paul Ohm, "Playing with the Data: What Legal Scholars Should Learn About Machine Learning" [2017] 51 *U.C. Davis L. Rev.* 654 – 717, 676; M. Hildebrandt, "Algorithmic regulation and the rule of law" (2018) 376 *Phil. Trans. R. Soc. A* 1-11, 7.
[22] Lehr and Ohm, "Playing with the Data," 665.
[23] Scantamburlo et al., "Machine Decisions and Human Consequences," 53.

This model-building process relies on several critical assumptions. First, that the training data serves as appropriate 'ground truth', a matter to which we will return. Secondly, that historic data provides a reliable guide to *future* instances of the phenomenon it is taken to represent: this assumes that the relevant practices and their surrounding context remain stable and unchanged, and that the training data was collected in a context relevantly similar to the proposed context of deployment: e.g., new forms of spam might not be accurately identified by a spam-predictor that has not included training data with that kind of spam. However, the conventional training of computer scientists might not extend to matters concerning the collection of training data and/or its relationship to real-world phenomenon which their models seek to predict beyond a narrow, relatively limited set of technical requirements. That said, the work of data scientists involves attending to the 'quality' of datasets to render them useful for algorithmic modelling, although data 'quality' is context dependent. Poor quality data includes 'dirty' data, suffering from problems that later hinder the quality of a given algorithmic model thus requiring human 'cleaning'[4] to render the data suitable for model-building,[24] or the data may be incomplete, noisy, contain significant outliers, infused with errors (caused by a variety of human and computer fallibilities, such as a failure in code, or a failure in human data entry) or formatted in a manner that is not processable by modelling software. As the European Fundamental Rights Agency (FRA) has observed, the use of 'low quality' data to produce prediction models that produce 'low quality' outcomes to inform real-world decision-making can lead to violations of fundamental rights, particularly rights to privacy and data protection[25] while negatively affecting the right to an effective remedy, lamenting the failure of many textbooks and articles dealing with data science to overlook these crucial aspects of data quality.[26] Identifying and selecting the appropriate input and training data requires *active* involvement and normative judgement by the data scientist to 'wrangle' a dataset into a useable format. However, even if technical developers regard themselves as professionally responsible for some matters of data quality, this might not extend to the legality of data collection and processing.[27] For algorithmic tools intended to assist criminal justice authorities, routinely collected administrative data is often used. Yet even assuming that such data can be lawfully collected, it does not necessarily follow that it can be lawfully used to build algorithmic tools[28] for at least two reasons:

(a) Unlawful data collection and processing

Firstly, privacy laws and contemporary data protection laws restrict the collection and processing of 'personal data' (i.e., data pertaining to an identified or identifiable individual) in significant ways.[29] Algorithmic tools used for criminal justice purposes have been found to violate these laws.[30] For example, the Hague District Court ruled that the state could lawfully create and employ data-driven technologies to identify fraudulent activity, SyRI was disproportionately intrusive and thus not legally justified as 'necessary in a democratic society', violating the Art 8(1) right to private life.[31] In a related vein, the UK Information Commissioner's Office (ICO) ruled that by naming identifiable individuals and labelling them 'Gang Nominals', the Gangs Matrix violated data protection laws. Although the data parsed by the Matrix to generate individual harm scores is drawn primarily from administrative data lawfully collected by the London MPS, it relied upon the excessive collection of personal data including 'gang association data' on informal unregulated lists, entailed unregulated data sharing between partner agencies, and included victim data in the Matrix without distinguishing them from suspected 'Gang Nominals.'[32]

---

[24] Tarleton Gillespie, 'The Relevance of Algorithms' in T. Gillespie, P.J. Boczkowski and K. Foot (eds.) *Media technologies: Essays on Communication, Materiality, and Society* (MIT Press, 2014), 171.

[25] European Union Agency for Fundamental Rights, "Data quality and artificial intelligence: mitigating bias and error to protect fundamental rights" (11 June 2019) *https://fra.europa.eu/sites/default/files/fra_uploads/fra-2019-data-quality-and-ai_en.pdf*.

[26] European Union Agency for Fundamental Rights, "Data quality and artificial intelligence," 3.

[27] Andrew Selbst, Suresh Venkatasubramanian and I. Elizabeth Jumar, "The legal construction of black boxes: How machine learning practice informs foreseeability" (2021) Paper presented at the We Robot 2021 conference, https://werobot2021.com/wp-content/uploads/2021/08/Kumar_et_al_Legal-Construction-of-Black-Boxes.pdf. (cited with the kind permission of the authors).

[28] It might be lawful to use some data train a model, but not to use the same set of features about a person to make/inform a decision about them using the model in the real world, or vice-versa. See ICO guidance on why controllers need distinct lawful bases for development and deployment: Information Commissioner's Office, "What do we need to do to ensure lawfulness, fairness, and transparency in AI systems?" at *https://ico.org.uk/for-organisations/guide-to-data-protection/key-dp-themes/guidance-on-ai-and-data-protection/what-do-we-need-to-do-to-ensure-lawfulness-fairness-and-transparency-in-ai-systems/?q=profiling*. We are grateful to Reuben Binns for drawing this to our attention.

[29] Article 4(1) of the EU's General Data Protection Regulation 2018 provides that 'personal data' applies only to information 'relating to an identified or identifiable natural person ('data subject').' See Case C-582/14 *Breyer v Germany* [2014] ECLI:EU:C:2016:779).

[30] Although there is room for debate about whether the resulting model could be considered unlawful *per se*.

(b) Increasing the likelihood of the unlawful exercise of public decision-making authority

Secondly, developers may regard the legal duties applicable to public authority decision-making as matters as outside their problem-space. But without a proper understanding of these duties, the tools they build may facilitate unlawful decision-making. Moreover, decisions attached to a particular office (including those taken by criminal justice officials, such as the police) are typically accompanied by specific legal duties, including administrative law duties that restrict the lawful exercise of discretionary power in the hands of public officials. For example, British administrative law requires that public power be exercised in accordance with the purpose for which that power was conferred, and only on the basis of 'relevant considerations,' while 'irrelevant' considerations must be excluded.[33] Accordingly, data used to train the algorithmic model must generate outputs that are legally 'relevant' to the decisions those outputs inform. But if the phenomena that the training data is taken to represent does *not* conform with legal requirements, then the resulting model may generate predictions that are legally 'irrelevant' to the decision at hand.[34] This might occur, for example, if the training data used is not a valid or lawful indicator of the predicted phenomenon. So, if legislation requires consider of variable A only, it is not legally permissible for the decision-maker to consider variable B.[35] Yet if no 'ground truth' data concerning the matter which a public authority is legally required to consider when making particular decisions is available, data scientists might utilise data-sets they regard as a 'proxy'.[36] This creates serious risks that the model's outputs will be legally irrelevant to the decision at hand resulting in decision-making that is ultra vires and therefore unlawful.

For example, consider a police officer in England or Wales who is legally required to decide whether an arrested person in custody (D) should be released without charge (or under continuing investigation), charged and publicly released on bail, or charged and detained in custody on remand prior to trial.[37] Continued detention on remand pending trial is only lawful if the conditions of s 3(6) of the Bail Act have been met and authorised. This requires establishing (among other things) that detention is deemed 'necessary' to secure D from absconding, to prevent D committing an offence while on bail, or to protect D or another person.[38] It is in making these kinds of determinations that US criminal justice authorities have enthusiastically embraced algorithmic tools that purport to assess the risk that an individual will abscond ('flight risk') or commit a criminal offence if released (so-called 'recidivism risk'),[39] although we are not aware of their use by British courts for these purposes. In assessing the necessity of retaining D in custody for the purposes permitted by the Act, some kind of assessment of the likelihood that the individual will commit a criminal offence, engage in self-harm, or harm another person during the period prior to trial is needed. To create an algorithmic model capable of providing such predictions, 'ground truth' data would consist of a comprehensive historic data set identifying the entire universe of crimes, and of individuals who have engaged in criminal conduct or otherwise harmful conduct directed at others or themselves for the geographic region in question over a substantial yet recent time-period. Yet because not all crimes or harmful actions are identified, reported, and recorded, and many of those who commit crimes and/or harms evade apprehension and conviction, in addition to a lack of systematic data concerning self-harming activities falling short of suicide, ground truth data about the commission of *crime and harm* across a population, even within a more narrowly defined geographic area, is not available and, indeed, is likely impossible to collect. This has not, however, prevented developers from building so-called 'recidivism risk' predictors using *arrest* data, treating arrest as a 'proxy' for crime committed. Arrest data is, however, an inadequate, highly misleading indicator of crime because not all arrested persons are charged, those who are charged may not be convicted, and many crimes are committed for which no arrests are made. Accordingly, the mere fact of arrest is *not* an acceptable proxy for the commission of criminal *offence*.

---

[31] ECHR Art.8.
[32] Information Commissioner's Office, "Enforcement Notice to the Commissioner of Police of the Metropolis" (13 November 2018), *https://www.met.police.uk/SysSiteAssets/media/downloads/force-content/met/about-us/gangs-violence-matrix/ico-enforcement-notice.pdf*.
[33] *R v Secretary of State for the Home Department, ex parte Venables and Thompson* [1998] AC 407
[34] Marion Oswald, 'Algorithm-assisted decision-making in the public sector: Framing the issues using administrative law rules governing discretionary power' [2018] 376 *Phil. Trans. R. Soc. A* 1-20, 10;
[35] The matter may be arguable if variable A can be consistently and reliably inferred from variable B. Difficulties can arise in evaluating the lawfulness of algorithmic models that have been generated through the use of datasets from which sensitive attributes (for example, such as gender or race) have been removed, because such sensitive attributes may well be readily inferred from other features in the data: see A. Roth and M. Kearns, *The Ethical Algorithm: The Science of Socially Aware Algorithm Design* (Oxford: OUP, 2019), 74-84.
[36] See discussion of the limitations of proxy data within the contexts of education, social care and criminal justice in Abigail Z. Jacobs and Hanna Wallach, "Measurement and Fairness" (2021) *FAccT' 21 Proceedings of the 2021 ACM Conference on Fairness, Accountability and Transparency*, 375-385.
[37] For a more expansive illustration of this decision process in the context of the use of algorithmic tools, along with applicable laws, see section 3.3. of Part I.
[38] Bail Act 1976, s.3(6): the decision whether to refuse bail and to remand D pending trial is made by a magistrate.

Thus, by failing to consider what the training data *actually* signifies, the tool's outputs may not offer meaningful indications of the phenomenon it purportedly predicts. Accordingly, the tool's outputs may be 'legally irrelevant' to the matter the public official must decide and therefore cannot lawfully be considered. In the above example, it would not be legally permissible to remand in custody an arrestee charged with a criminal offence solely on the basis of an algorithmic tool that predicts the likelihood that D will be *re-arrested* because the legal significance of a lawful arrest differs very significantly from that of a criminal *conviction*.[40] But it is on the basis of arrest data, for example, that the HART tool was developed, which is described as generating predictions about whether an individual is likely to 'commit a criminal offence' within a two year period. At the very least, these tools should be properly described as '*arrest* predictors' rather than mislabelling and misrepresenting them as predictions of future criminal offending.[41]

*3.2     Step 2: Apply a machine learning algorithm to create a prediction model*

Armed with labelled 'ground truth' data, ML software is then applied to train and build an algorithmic model (a 'classifier') that assigns a class label (e.g., 'spam or ham', 'diabetic or healthy', 'safe or dangerous') to unseen data. Advanced ML techniques enable the generation of more powerful and sophisticated mathematical models by analysing larger, more complex, highly dimensional data. Because many ML methods are available, technical developers must choose between them (and may experiment with several alternatives) including regression models that generate a score for each candidate or classification models that allocate candidates into classes. One parameter of great importance, particularly for algorithmic tools used in rights-critical contexts, concerns the configuration of the model's 'error thresholds.'[42] Predictions generated using ML models are inherently probabilistic: prediction errors are therefore unavoidable. In this context, errors are defined as the failure of the model to generate an accurate prediction on unseen data, either in the form of a Type I error (false positive) or Type II (false negative).[43] Good data scientists also recognise the vital importance of attending to the distribution of Type I and Type II errors in building algorithmic models, requiring careful consideration of their real-world consequences. For example, consider the consequences of errors produced by an algorithmic tool that evaluates skin melanoma images to predict whether a melanoma is cancerous or benign, intended for at-home diagnosis to help individuals decide whether to seek medical advice. A Type II error (false negative) is considerably more serious: a person who mistakenly believes that the melanoma is benign may not seek further medical treatment, allowing the cancer to develop untreated, with potentially fatal consequences. In contrast, a Type I error (false positive) is likely to prompt that person to seek medical assistance unnecessarily, assuming that the melanoma is subsequently diagnosed correctly by the clinician as benign.[44]

---

[39] Kathrin Hartmann & Georg Wenzelburger, "Uncertainty, risk and the use of algorithms in policy decisions: a case study on criminal justice in the USA" (2021) 54 *Policy Sciences* 269-287.
[40] Virginia Eubanks, *Automating Inequality: How High-Tech Tools Profile, Police and Punish the Poor* (New York: St. Martin's Press, 2018). David Robinson "The Challenges of Prediction: Lessons from Criminal Justice." (2017) 14 *Journal of Law & Policy for the Information Society Robinson,* 151.
[41] Robinson, "The Challenges of Prediction," 160-161.
[42] Steven W. Knox, *Machine Learning: A concise introduction* (Chichester: Wiley, 2018), 15-32; D. Spiegelhalter, *The Art of Statistics: Learning from Data* (Pelican, 2020).
[43] Spiegelhalter, *The Art of Statistics: Learning from Data*.

Careful consideration of the real-world consequences of different error-types is particularly important when determining the cut-off point at which point the model assigns a classification. As Scantamburlo et al explain, when ML techniques are used to build binary classifiers that can assign an item to one of many possible categories (such as 'spam' or 'ham'), many (although not all) involve in a two-stage process: first, a real valued score is computed for the item to be classified, and secondly, that score is compared with a cut-off point whereupon the item is assigned to a class depending on whether it exceeds the cut-off point.[45] The real valued score could informally be thought of as a probability, though it is not necessarily a formal probability. In some (but not all) cases, changing the threshold results in a trade-off between false positives (Type I error) and false negatives (Type II error). In the example of skin-cancer predictor, this can be captured by notions of 'sensitivity or recall' (i.e., the true positive rate is the proportion of true positives in relation to the aggregate of true positives + false negatives) and 'specificity' (i.e., the true negative rate is the proportion of true negatives in relation to the aggregate of true negatives + false positives). For example, for our melanoma classifier, because the consequences of mistakenly classifying a melanoma as benign when it is in fact cancerous (Type II error) is far more serious than mistakenly classifying a benign melanoma as cancerous (Type I error), the cut-off point should be configured to minimise Type II errors, even at the cost of increasing Type I errors. These examples also demonstrate that the real-world consequences of error are context-dependant: a Type I error in 'recidivism risk' prediction is far more serious for an affected individual than a Type I error by an automated spam-filter.

As a matter of constitutional principle, algorithmic tools that inform how an individual will be treated by criminal justice authorities must be configured to distribute the risk of error in a manner that respects due process rights, including the presumption of innocence protected under ECHR Art 6. These procedural rights are rooted in the state's duty to demonstrate respect for persons. It demands that those adversely affected by the decision of a public authority, particularly criminal justice authorities wielding the coercive power of the state, are entitled to be informed of, and to challenge those decisions, particularly given the ever-present danger of mistakes in decision-making. As a matter of administrative law, the specific procedural requirements (and concomitant duties) that the right to due process demands is highly context-dependent, in which the decision's impact upon the affected individual's rights, interests and legitimate expectations has great importance.[46] Hence a person whose driver's licence application is refused enjoys a fairly limited set of procedural rights,[47] while a person charged with committing a very serious criminal offence is entitled to an extensive suite of procedural rights, typically including rights to free legal representation, to the application of strict rules of evidence and procedure, and to the presumption of innocence in which the burden of providing the accused's guilt 'beyond reasonable doubt' is placed firmly on the prosecution.[48] This high standard of proof, which is far more demanding than the 'balance of probabilities' standard applicable to civil cases, produces a legal system systematically designed to minimise Type I errors (false positives) while producing more Type II errors (false negatives). This means guilty defendants may avoid conviction because the prosecution has failed to prove guilt 'beyond reasonable doubt.' Although such outcomes are deeply regrettable, these errors are nevertheless widely accepted in modern western European legal systems as worth incurring to avoid the far more egregious injustice associated with Type I errors—that is, wrongfully convicting an innocent person. Particular care is needed for prediction models created by ML techniques which exacerbate the dangers associated with mistakes because they are produced on the basis of correlations in the underlying training data rather than on scientifically established causal relationships.[49]

---

[44] I.e., the patient will not then be subjected as very invasive, unnecessary interventions due to an incorrect diagnosis. See for example Karsten Juhl Jørgensen, Peter C. Gøtzsche, Mette Kalager, and Per-Henrik Zahl. "Breast Cancer Screening in Denmark: A Cohort Study of Tumor Size and Overdiagnosis" [2017] 166(5) *Ann Intern. Med.* 313-323.
[45] Scantamburlo et al., "Machine Decisions and Human Consequences," 54.
[46] D. Galligan, *Due Process and Fair Procedures: A Study of Administrative Procedures* (Oxford: OUP, 1996).
[47] Including rights to be tested by an unbiased examiner, to be given a reasonable opportunity to demonstrate that their driving skills meet the requisite legal standard, and to be provided with reasons for any refusal of a license.
[48] ECHR, Art. 6.

To demonstrate how human rights should condition algorithmic tool-design, consider again the decision to detain or release a person charged with a criminal offence (D) pending trial. To detain a person against her will for extended periods entails a grave and serious denial of liberty. If these decisions are to be informed by predictions that purport to classify an individual as 'dangerous' or 'harmless' such that the former is held in remand while the latter released, then falsely classifying a harmless person as 'dangerous' (Type I error) will entail a serious interference with that person's right to liberty under Article 5 ECHR.[50] However, falsely classifying as 'harmless' a person ('D'), who is in fact dangerous, will result in release, placing the public at risk that D will commit a serious crime during the pre-trial period. Nevertheless, if we are to take the presumption of innocence and the individual's right to liberty seriously, this necessitates treating false positives (convicting the innocent) as far more serious than false negatives (failing to convict the guilty). As a matter of constitutional principle, this relative weighting must be reflected in the construction of the underlying mathematical model, although the human decision-maker might justifiably override the algorithmic recommendation if release may expose specific, vulnerable individuals to a substantial risk of domestic violence. Although this means the public systematically bears a proportionately greater risk that a dangerous person will be released pending trial, constitutional democratic societies accept that this is the inescapable cost of upholding individual rights and freedoms. If the model were to be configured otherwise, this may unlawfully violate the arrestee's Article 5 ECHR right to liberty and security, including freedom from arbitrary detention, and will also increase the likelihood of further violation of her Article 6 ECHR rights, including the right to be presumed innocent and the right to a fair trial.

Despite the importance of these technical choices, there is little public information concerning how the risks of Type I and II error are configured into algorithmic tools used by criminal justice authorities. Of our three case studies, public information was only available in relation to the configuration of error thresholds for the HART tool, thanks to Urwin's published Master's thesis and the account provided in a jointly authored paper with Barnes, Oswald and Grace.[51] Yet those accounts reveal a disturbing failure to demonstrate any recognition that an arrested person is entitled to be presumed innocent, stating that error thresholds were configured on the basis that "it is worse to misclassify a dangerous offender as harmless than to erroneously classify a harmless individual as dangerous."[52] Although this may reflect the perspective favoured by the general public, it fails to recognise that such a choice entails an unjustified violation of the affected individual's right to be presumed innocent (and hence harmless).[53]

*3.3  Step 3: Model testing and validation*

Having generated a mathematical model, computer scientists then conventionally assess its 'accuracy,' understood as how well it correctly predicts outcomes from a set of unseen historic data. Binary classification algorithms of the kind under consideration are conventionally and primarily evaluated in terms of accuracy in predicting the correct classification of candidates taken from the 'test set' (or 'out of bag sample') namely data held back and kept separate from the remainder of the dataset used to train and tune the model.[54] Accuracy in this context conventionally refers to the percentage of accurate predictions calculated as a ratio of the total number of correct predictions (that is, the total number of true positives and false negatives) relative to the total number of predictions generated by the model.[55] But there are many other commonly used quality metrics that can be employed to evaluate a tool's 'accuracy' with Krafft and Zweig[56] identifying 31 different and 'frequently used' quality measures. These include, for example, precision and recall, which we have already considered[57], as well as the AUC-ROC curve.[58] They argue that developers should select quality measures that are most suited to the social and organisational context in which the algorithm is to be used, yet they observe that this is not always the case.[59] Furthermore, these mathematical assessments of 'accuracy' typically overlook the fact that the datasets themselves are of dubious validity as a basis for predicting what an individual will do if publicly released: although they will include data on whether a person who was released was subsequently arrested, they will not contain data about whether those detained in custody *would have* committed a crime had they been released.

---
[49] See section 3.1 of Part I for further discussion of the constitutional acceptability of using ML models to make predictions about individuals in rights-critical contexts.
[50] ECHR, Art. 5.
[51] Oswald et al, in "Algorithmic risk assessment policing models," state at 236: 'The HART model represents a real example of a value-judgement built into an algorithm, so requiring a 'trade-off' to be made between false positives and false negatives in order to avoid errors that are thought to be the most dangerous: in this context, offenders who are predicted to be relatively safe, but then go on to commit a serious violent offence (high risk false negatives). As a consequence, high risk false positives have been deliberately made more likely to result.'
[52] Urwin, "Algorithmic forecasting of offender dangerousness."
[53] Further, HART's error thresholds were unjustified because they were insufficiently tailored to the policy purpose for which it was introduced to serve. Neither high risk false positives, where the individual was actually low risk, nor high false negatives where the individual was actually high risk, result in any change in the substantive recommendation. High and low risk scores could not grant entry into Checkpoint. Only if a person who is medium risk is wrongly classified as high or low risk would they be erroneously denied entry. If individuals who are low or high risk were erroneously classed as medium risk, they would then be granted entry, when in fact the policy intention was to exclude them. On the presumption of harmlessness, see A. Ashworth and L. Zedner, *Preventive Justice* (Oxford: OUP, 2014), 53.

There is, however, increasing recognition by data scientists of the need to evaluate 'quality' from values other than accuracy in prediction. Those working within the field of fair-ML or 'FAccT' computing are actively investigating how ML models may be evaluated by reference to values such as 'fairness,' 'explainability' and 'transparency.'[60] Yet these approaches typically adopt a very narrow and somewhat contrived understanding of normative values, conceived largely in mathematical terms which then lend themselves to quantification and computational analysis.[61] This is particularly true of 'fairness', with researchers seeking to define aspects of inherently vague notions of societal fairness in mathematical terms in order to incorporate fairness ideals into machine learning.[62] However, as critics observe, these definitions are simplifications that fail to capture the full range of similar and overlapping notions of fairness and discrimination in philosophical, legal and sociological contexts.[63]

Apart from the dangers associated with relying upon ML-based prediction models for which a causal understanding of the variables remains unknown (examined in Part I), these quality metrics focus exclusively on the algorithmic tool rather than the larger socio-technical system in which they operate producing a 'framing trap' that leads to inappropriate and misleading 'quality' guarantees.[64] We argue that algorithmic tool 'quality' also demands adherence to the basic constitutional principles and legal requirements, particularly when used to inform 'rights-critical' decisions by public authorities constituting constitute vitally important and relevant socio-technical context. Hence the 'quality' of these tools should be assessed as constitutionally and legally acceptable before they can be used to inform real-world decisions. These assessments require careful, context-sensitive, qualitative legal evaluation that cannot be collapsed or reduced to purely mathematical terms. To do so risks overlooking rights-critical normative choices made by developers in the design process of algorithmic tools, enhancing the dangers of illegality and injustice.

---

[54] D. Michie, D. Spiegelhalter and C. Taylor (eds.) *Machine Learning, Neural and Statistical Classification* (New Jersey: Prentice Hall, 1995), 15. A 'validation set' refers to at least one portion of data from the training dataset conventionally set aside to validate the model and tune its parameters.

[55] Kaitlin Kirasich, Trace Smith and Bivin Sadler, "Random forest vs logistic regression: Binary classification for heterogenous datasets" (2018) 1(3) *SMU Data Science Review* 1-24, 12-14.

[56] Tobias D Krafft and Katharina A Zweig "So far, So Good: Multidisciplinary perspectives on algorithms, decisions and algorithmic decision-making – Computer Science Dimensions." (2019) *Unpublished manuscript*.

[57] See section 3.2.

[58] Area under the curve – receiver operating characteristic: Melissa Hamilton "Adventures in risk: predicting violent and sexual recidivism in sentencing law" [2015] *Arizona State Law Journal* 47(1) 11-62, 34-37.

[59] E.g., demonstrating that the purported AUC-ROC score of COMPAS (an algorithmic tool developed by Northpointe Inc. to undertake 'recidivism risk' assessment and inform sentencing decisions), offers very little meaningful information about the tool's accuracy in, nor suitability for, this purpose: Krafft and Zweig 'So far, So Good…' at 8.

[60] e.g., see the ACM Conference on Fairness, Accountability, and Transparency (ACM FAccT): https://facctconference.org/; Roth and Kearns, *The Ethical Algorithm: The Science of Socially Aware Algorithm Design*.

[61] C. Barabas, "Beyond bias: 'Ethical AI' in criminal law" in M. Dubber, F. Pasquale and S. Das (eds.), *The Oxford Handbook of Ethics of AI* (OUP, 2020).

[62] See the well-known debate about the fairness of COMPAS risk scoring: Julia Angwin, Jeff Larson, Surya Mattu, and Lauren Kirchner, 'Machine bias: There's software used across the country to predict future criminals. And it's biased against blacks" (ProPublica, 2016) at *https://www.propublica.org/article/machine-bias-risk-assessments-in-criminal-sentencing*; William Dietrich, Christina Mendoza and Tim Brennan, "COMPAS risk scales: Demonstrating accuracy equity and predictive parity (2016)" Northpointe Inc. *https://go.volarisgroup.com/rs/430-MBX-989/images/ProPublica_Commentary_Final_070616.pdf*.

[63] Ben Green, "The false promise of risk assessments: Epistemic reform and the limits of fairness" (2020) *FAT*'20: Proceedings of the 2020 Conference on Fairness, Accountability and Transparency*, 594-606; Roth and M. Kearns, *The Ethical Algorithm: The Science of Socially Aware Algorithm Design*, Ch.2.

*3.4     Step 4: Develop a digital interface for use in a particular organisational setting*

Armed with a suitable prediction model, a digital user-interface is then designed to communicate the model's predictions to human decision-makers. The widespread take-up of smart devices substantially enhances the ease and convenience with which front-line officers can access algorithmic predictions. Although convenience, efficiency and seamlessness are understandably important from a user-design perspective, there is a worrying tendency to downplay or overlook the need to design interfaces in ways that respect the due process rights of affected persons. We have already argued that respect for due process rights should be demonstrated when configuring error thresholds within algorithmic models. This is also true when designing digital user-interfaces for front-line decision-makers. Although identifying the specific set of procedural requirements (and concomitant duties) that the right to due process imposes requires is only possible in specific application contexts, three matters are especially important yet often overlooked by UX designers trained to prioritise organisational convenience and efficiency, briefly outlined below:

i. **The right to reasons**: under British Administrative law, an individual's right to reasons for public authority decisions that have significant adverse effects flows from the legal duty of public authorities to explain and justify their decisions in accordance with law.[65] If an algorithmic tool produces a score and/or recommendation about an individual without an accompanying explanation of what it is intended to signify, or how it was generated, the front-line official receiving it may not be able to provide the affected individual with reasons to support it. Accordingly, user-interfaces for algorithmic tools to support public authority decision-making should, as a matter of constitutional best practice, if not of legal obligation, provide functional explanations to accompany the tool's output(s) (at least for 'interpretable' ML models). Front-line officers should also be able to identify and cross-check how the output was produced, lest over-simplified explanations exacerbate the likelihood of automation bias, discussed below.[66] Similarly, tool outputs must not be mislabelled: a tool intended to predict 'recidivism risk' developed from arrest data should *not* be called a 'recidivism' predictor, but a 're-arrest' predictor. The legal duty of public authorities to give reasons for decisions that entail the adverse treatment of individuals suggests that they should not employ decision-support tools that use advanced ML techniques such as deep neural networks, for which even functional explanations remain elusive.[67] Yet, for the three case studies examined here, we could not identify whether front-line decision-makers were automatically provided with explanations to accompany the algorithmic predictions intended to assist them. Much more extensive and systematic transparency regarding decision-support tools used in criminal justice is therefore needed.[68]

ii. **The right to contest:** providing explanations about the tool's functional logic, including the relevant variables upon which it relies, might help decision-makers discharge their administrative law duty to provide affected individuals with 'reasons' for a particular decision in specific cases. However, those reasons must be *lawful* reasons[69] and good faith decision-making is not sufficient.  Hence the individual's right to challenge and contest such decisions has considerable constitutional importance, reflected in rights to liberty, to due process and to a fair trial protected under ECHR Articles 5 and 6. While the Data Protection Act 2018 (and the GDPR and Law Enforcement Directive upon which the 2018 Act builds) confers on data subjects a right to contest fully automated decisions,[70] contestation rights must also accrue to those significantly adversely affected by a recommendation produced by an algorithmic tool to guard against injustice and the abuse of power.

---

[64] Selbst et al., "The legal construction of black boxes."
[65] *R v Secretary of State for the Home Department, Ex parte Doody* [1993] UKHL 8; Jennifer Cobbe, "Administrative Law and the Machines of Government: Judicial Review of Automated Public-Sector Decision-Making" 39(4) *Legal Studies* 63.
[66] We are indebted to Mireille Hildebrandt for this point.
[67] See Rebecca Williams, "Rethinking administrative law for algorithmic decision-making" (2021) 42(2) *Oxford Journal of Legal Studies* 468-494, 482.
[68] See section 5.
[69] Williams "Rethinking administrative law for algorithmic decision-making," 482.

iii. **The right to an unbiased tribunal**: Although we have emphasised the 'fair hearing' limb of the administrative law right to procedural fairness, its second limb - the 'rule against bias' - is equally important. It requires that decision-making tribunals must be 'impartial', meaning free from both actual bias and the *appearance* of bias.[71] The issue of bias and discrimination, particularly racial and gender bias of algorithmic tools used in criminal justice decision-support, has received widespread attention.[72] Datasets inevitably reflect underlying biases in the historic social practices to which the data pertains so that, if used to generate prediction models, the resulting outputs will reflect and reinforce these biases. Historically marginalised groups are thus subjected to a higher risk of unjust discrimination relative to individuals from majority groups because the resulting outputs *systematically* discriminate unjustly (rather than being arbitrary, they are systematically patterned for reasons we can point to) and may violate the right to be free from unjust discrimination in the determination of opportunities and burdens. To this end, the ICO concluded that the London Gangs Matrix was unlawful because the MPS failed to ensure that its use complied with the Public Sector Equality Duty (s.149 of the Equality Act 2010) because it disproportionately singles out black men relative to other ethnic groups.[73]

Less attention, however, has been paid to automation bias,[74] referring to the human tendency to trust machine-made judgments over their own despite their potential or demonstrated capacity for error. Although algorithmic 'recommender' tools may formally preserve human judgement, front-line officials may in practice tend to follow recommendations unreflectively.[75] As Johnson and Powers[76] commented in the context of computerised aviation systems in which a human 'in-the-loop' is expected to supervise computational systems, those individuals may be understandably reluctant to intervene.[77] Similarly, front-line criminal justice decision-makers under considerable time pressures, heavy workloads and lack a clear understanding of how algorithmic recommendations are produced are unlikely to depart from them. Accordingly, officials who use algorithmic tools must be properly trained so that they can properly understand and interpret algorithmic recommendations and their limitations, and to ensure they exercise meaningful independent judgement rather than unthinkingly following the tool's outputs.

**4.       Discussion and Recommendations for Reform**

We have demonstrated how important constitutional principles and legal duties are implicated at every step of the algorithmic model-building process for use by criminal justice decision-makers yet largely overlooked in relation to three such tools used to date. As a result, these tools may significantly enhance the risk that decisions based on their predictions may be unjust or may otherwise entail the unlawful exercise of decision-making authority. To minimise these dangers, it is vital that during the algorithmic tool-building process, proper consideration is given to:

**(a)** the nature of the *substantive interventions* that flow the outputs generated by these prediction tools, particularly effects on the rights, interests, and legitimate expectations of those subjected to algorithmic evaluation;

---

[70] Williams "Rethinking administrative law for algorithmic decision-making ," 474-476.
[71] *Ridge v Baldwin* [1964] AC 40.
[72] Cathy O'Neil, *Weapons of Math Destruction* (New York: Crown Books, 2016).
[73] Information Commissioner's Office, "Enforcement Notice to the Commissioner of Police of the Metropolis" at [41].
[74] Linda J. Skitka, Kathleen L. Mosier, Mark Burdick, and Bonnie Rosenblatt, "Automation bias and errors: Are crews better than individuals?" (2000) 10(1) *The International Journal of Aviation Psychology* 85-97 (2000); ICO (n 101) at [41].
[75] Hartmann & Wenzelburger, "Uncertainty, risk and the use of algorithms in policy decisions," 269-287.
[76] Deborah G. Johnson and Thomas M. Powers, "Computer systems and responsibility: A normative look at technological complexity" (2005) 7 *Ethics and Information Technology* 99-107, 106.
[77] Arthur Kuflik, "Computers in control: Rational transfer of authority or irresponsible abdication of autonomy?" (1999) 1 *Ethics and Information Technology* 173-184.

**(b)** the *legal duties and obligations* that apply to all public officeholders who make decisions about the treatment of individuals within the criminal justice process;

**(c)** the *public policy objectives* that the larger criminal justice system in which the tool is embedded is intended to serve, and

**(d)** the need to maintain the general *public's trust and confidence* in the integrity of the criminal justice system and the administration of justice.

Yet, as we argued in Part I, computer scientists are conventionally trained to abstract or 'detach' the prediction model from legally and constitutionally relevant considerations. By focusing exclusively on developing mathematical models to generate 'accurate' predictions understood in narrow, technical terms, the resulting tool may violate legal requirements and constitutional principles, resulting in decisions that entail the unlawful or improper exercise of public power. Despite the intuitive, common-sense need to consider real-world consequences and legal concerns when approaching the task of algorithmic tool-creation, they tend to be ignored by technical experts who may conventionally regard them as outside the relevant 'problem space' and hence 'not their responsibility.' It is therefore essential that technical experts work collaboratively with legal professionals with a strong grasp of the appropriate constitutional principles, human rights norms and doctrines of administrative law when creating, testing, and evaluating algorithmic tools prior to their deployment by public decision-makers.[78] However, because these interdisciplinary teams may want more tractable, concrete guidance to better understand how constitutional principles should apply to, and be 'operationalised', in the tool-building process, we offer brief advice, focusing on three sets of 'detachment' practices that should be steadfastly *avoided*.

*4.1.    Detaching the tool from the substantive interventions that its predictions seek to inform*

We have shown that algorithmic tools cannot legitimately be designed, evaluated nor deployed without due consideration of their real-world consequences**,** requiring careful attention to the *specific context* of their deployment and application domain. Yet technical specialists conventionally focus on the accuracy of algorithmic predictions, without taking due account of the effects of the substantive interventions which follow. Thus, the consequences of *mistaken* outputs upon individuals subjected to algorithmic evaluation are not given due consideration. Although mistaken on-line consumer product recommendations may be inconsequential to the recipient, this is not true of mistaken, biased, or spurious algorithmic predictions that inform rights-critical criminal justice decisions, particularly about coercive detention. Throughout the tool-building process, attention must be paid to the substantive interventions intended to follow from predictions thereby generated, particularly their resulting impact on the affected individual, to ensure that:

- The model's error thresholds appropriately reflect the due process rights of the affected individual, including the right to be presumed innocence (and the presumption of harmlessness) where appropriate;

- Due consideration is given to the normative acceptability of utilising the algorithmic models to generate predictions to inform decisions that may adversely affect the individual under evaluation, particularly in the absence of scientific evidence of causal relations upon which the tool relies. Algorithmic tools should not be deployed in rights-critical contexts unless a plausible account of why the features upon which the model relies to generate productions may be expected to provide reliable and truthful indicators of the phenomenon which the model seeks to predict[79]; and

- The design of the algorithmic tool's user-interface for communicating outputs to front-line officers must enable them to understand what the output signifies—including its limitations—and remind the officeholder of the rights of affected persons to demand reasons for any significant adverse decision, to contest the decision, to an unbiased tribunal, and to help the officer ensure that meaningful, independent judgement is brought to bear when making the resulting decision.

---

[78] Michael Veale, Max Van Kleek and Reuben Binns, 'Fairness and accountability design needs for algorithmic support in high-stakes public sector decision-making' [2018] *Proceedings of the 2018 CHI Conference on Human Factors in Computing Systems* 440.
[79] For further discussion, see section 3.1. of Part I.

*4.2     Detaching the tool from its surrounding legal context and/or underlying policy purpose*

Failing to properly attend to the surrounding legal and policy context during the design and construction of algorithmic decision-support tools enhances the likelihood that these tools will produce predictions that give rise to unlawful decision-making and may generate serious injustice. Accordingly, when building such tools, due care must be given to:

- particular matters which the law requires the decision-maker to decide, including specific considerations mandated by law that must be considered in decision-making. For example, if the statute requires consideration of the likelihood of an individual committing a crime in future, then it should not be assumed that this is equivalent to the likelihood of being *arrested* in future;

- ensuring that only 'legally relevant' considerations ('features') are analysed in the model generation process, and that irrelevant features are disregarded;

- attending to the congruence between the phenomenon that the tool is intended to predict and the substantive decisions which the public official is legally obliged to make. Thus, if the decision-maker is required to consider the 'dangerousness' of an individual, predictions about the likelihood of being arrested in future may not be reliable indicia of dangerousness; and

- careful evaluation of whether the proposed ground truth data constitutes a valid and acceptable proxy for generating predictions about the phenomenon of interest.

*4.3     The need for systematic safeguards: attending to the power, opacity, and scale of algorithmic tools*

We have highlighted how algorithmic risk assessment tools used for criminal justice purposes may result in the unjust and/or unlawful treatment of individuals. But transparency and accountability must also be provided to the public at large, otherwise there is no basis for them to trust that governmental power is not being abused, exploited, or otherwise exercised corruptly. Accordingly, the exercise of governmental authority—particularly in criminal justice contexts—must be subject to systematic, institutional oversight, to ensure that this authority is exercised in a lawful and accountable manner, including mechanisms to ensure that the exercise of that authority in individual cases is subject to meaningful contestation, review, and redress. A commitment to constitutionalism demands that endowed with governmental power exercise it in an open and transparent manner, routinely rendering an account for the exercise of their decision-making power *to the community at large* from whom their power is ultimately derived and on whose behalf they purport to act. Yet relatively little attention is given to the opacity and power of these tools arising from their scale and speed of operation. Instead, we observe a third, equally if not more troubling form of 'detachment', generating serious constitutional dangers: the inaposite use of analogical reasoning evident in some policy-making and judicial reasoning, whereby advanced data-driven tools are detached from their capacity to be employed automatically and at scale. Instead, algorithmic tools are often treated as 'equivalent' to existing (analogue) tools long used to inform criminal justice decisions, on the basis that they pursue the same purposes. This 'fallacy of equivalence' is used in at least two ways in legal analysis that are both mistaken and dangerous.[80]

Firstly, it is constitutionally mistaken to assume that the legal basis authorising the use of an 'old fashioned,' handcrafted, statistical tool also provides the legal basis for a data-driven algorithmic tool capable of being automated at scale, merely because the latter serves the 'equivalent' purpose. This *inappropriate invocation of the argument from analogy* overlooks the massively enhanced power of data-driven technologies, their opacity, the susceptibility of humans to automation bias, and the instant reproducibility, transfer and storage of digital data.[81] For example, in a judicial review challenge to the South Wales Police use of live facial recognition technology (FRT), the High Court reasoned that having one's face subjected to analysis by live FRT was no different from having one's face photographed by an police officer in the course of his or her duty, which earlier judicial decisions had confirmed may be lawfully undertaken by police on the basis of their common law powers to prevent and detect crimes.[82] Such reasoning reflects a failure to understand that the technological capacities of contemporary data-driven, networked technologies are *qualitatively different* from an earlier generation of pre-internet enabled tools, used for the same or similar purposes. The danger here is that the false assertion of conformity may undermine the *rule of law,* while bypassing the need for public debate and deliberation about whether the general public considers it appropriate for these technologies to be employed by law enforcement authorities, and if so, on what terms and with what safeguards. Accordingly, we argue that the capabilities and limitations of new technological tools in real-world contexts must be properly considered when scrutinising claims that existing warrants of authority provide an adequate legal basis to authorise their use. If they expand and extend dangers associated with their use, *express* statutory authorisation must first be provided, facilitating open public debate and scrutiny rather than acquiesce in the expansion of the state's coercive powers by technological stealth.

---

[80] Karen Yeung, "Constitutional Principles in a Networked Digital Society" (2022), *https://ssrn.com/abstract=4049141*.

Secondly, the capacity of decision-support tools to operate automatically, at scale, yet in a highly opaque manner, must be taken into consideration when evaluating the 'proportionality' of deploying any tool, even assuming that it serves a legitimate and lawful purpose. We applaud the reasoning of the court in *SyRI* when concluding that, although state authorities could legitimately employ data-driven tools to identify welfare fraud, the gathering of administrative data from a wide range of unconnected sources to create highly detailed profiles of individuals was disproportionate, violating the ECHR Article 8(1) right to privacy.[83] In so doing, the Court provided important judicial acknowledgement that analogue tools cannot be considered 'equivalent' to their networked data-driven digital counterparts, even when employed to serve familiar and legitimate legal and policy objectives, owing to qualitative differences arising from their capacity to scale, and the level of intrusiveness that systematic data-collection across disparate contexts entails.

Although algorithmic tools offer organisations the possibility of automating tasks at scale, this radically magnifies their power and, in turn, the scope at which injustice and the abuse of power that they may generate. Indeed, the recent UK Post Office Horizon scandal in which defective financial accounting software produced by Fujitsu resulted in the wrongful conviction of 39 innocent individuals, considered the largest miscarriage of justice in British history, powerfully illustrates how software can radically 'scale injustice' in a systemic yet highly opaque manner which individuals may find practically impossible to contest.[84] It highlights the urgent need for systematic transparency, accountability and independent oversight to protect those subjected to algorithmic evaluation within the criminal justice system, which our analysis reveals are sorely lacking. In particular, institutional mechanisms of transparency and accountability are necessary in the decision to *adopt* these tools in the first place, and the way in they are built, deployed, and evaluated. The UK currently has no comprehensive, publicly available, inventory of decision-support tools used by criminal justice authorities that provides even basic information about such tools, let alone supplying evidence of their beneficial and adverse impacts. Indeed, there may be tools in use of which the public has no knowledge.[85] Without meaningful systematic oversight, algorithmic tools will continue being built and used in ways that automate and scale injustice, enhancing the risks that governmental power will be abused. Systematic independent oversight is urgently needed, including the creation and maintenance of a transparency register to help ensure that the safeguards we have identified have been properly implemented.[86] Such a register should, at minimum, include public information about the matters identified in section 4.2(d), as part and parcel of the 'right to reasons' including:

1. The legal authority upon which the tool may lawfully be developed and used, including the particular decisions that the tool is intended to inform;
2. The configuration of error thresholds and other normative trade-offs made during tool-development;
3. Information concerning how those normative choices were made and by whom;
4. Input, test, and validation data used to produce the underlying mathematical model, and upon which the tool relies to generate predictions;
5. Basic information about the outputs produced and what those outputs are taken to signify;
6. The type of machine learning used and why;
7. Measures concerning the tool-validation process employed, including relevant quality measures (including accuracy, recall and sensitivity);
8. The extent to which the model's predictions are supported by evidence demonstrating the causal underpinnings of the claimed relationship between the features used and the phenomenon which the model seeks to predict (or at least reasonable plausibility between those relations);
9. The intended 'users' of the tool, and the particular organisational and legal contexts in which they are intended to be used; and
10. The range of specific procedural and other safeguards in place to guard against decisions taken on the basis of error, illegality, irrationality or other forms of abuse of power, both individual and systematic.

---

[81] Skitka et al, "Automation bias and errors: Are crews better than individuals?"
[82] *R (Bridges) v South Wales Police and ors* [2020] EWCA Civ 1058 at [54]-[85].
[83] SyRI Judgment at [6.78]-[6.85].
[84] Haroon Siddique, "Wrongly convicted Post Office workers to get up to £100,000 interim payouts" (*The Guardian*, 22 July 2021), https://www.theguardian.com/business/2021/jul/22/wrongly-convicted-post-office-workers-to-get-up-to-100000-interim-payouts.
[85] House of Lords Justice and Home Affairs Committee, "Technology rules?" 40: "On the most basic level, we cannot be certain what technologies are being used for the application of the law in England and Wales."
[86] Also recommended by the House of Lords Justice and Home Affairs Committee, "Technology Rules?" at pp 45-46, relating to "all advanced algorithms used in the application of the law that have direct or indirect implications for individuals."

## 5. Conclusion

This two-part paper has demonstrated how seemingly 'technical' choices made by developers when building algorithmic tools for criminal justice authorities have serious constitutional implications which cannot be reduced to issues of technical computational know-how. We have argued that, because technical developers cannot reasonably be expected to have a proper understanding of public law principles, they must collaborate closely with public law experts when deciding whether to employ decision-support tools for specific criminal justice purposes, and if justified, to ensure they are configured in a manner that is demonstrably compliant with public law principles and doctrine, including respect for human rights, throughout the tool-building process. It is widely recognised that public law principles and safeguards apply to inform and constrain the use of weapons and other enforcement tools and techniques employed by the police and other criminal justice authorities, such as handcuffs, guns, tasers, teargas through to heightened surveillance, electronic tagging, and so forth. Yet there has been a peculiar ongoing failure to recognise that these same principles and safeguards should be applied to constrain and inform algorithmic decision-tools. Although public law scholars have highlighted how algorithmic tools may be in tension with constitutional values, they have been surprisingly reluctant to argue, and to demonstrate why constitutional principles operate as 'red lines' that mark out the boundaries of acceptability. In contrast, by employing cross-disciplinary insights from public law and data science, we have sought to demonstrate that unless and until these tools are designed and deployed in compliance with basic constitutional principles and legal requirements, they should not be used.[87] Systematic institutional oversight mechanisms are urgently needed to ensure such compliance, otherwise algorithmic tools are likely to proliferate in ways that violate individual rights, producing injustice and eroding public trust in the integrity of the criminal justice system.

**11.10.22**
**12158 words**

---

[87] Tobias D. Krafft, Katharina A. Zweig and Pascal D. König, "How to regulate algorithmic decision-making: A framework of regulatory requirements for different applications" (2020) *Regulation and Governance*, 1-18.